\documentclass[longbibliography,twocolumn,amsmath,amssymb,aps,nofootinbib]{revtex4-2}

\usepackage{graphicx}
\usepackage{natbib}
\usepackage{amsmath}
\usepackage{longtable}
\usepackage{gensymb}
\usepackage{enumerate}
\usepackage{varwidth}
\usepackage{float}
\usepackage{nonfloat}
\usepackage{soul}
\usepackage{ulem}
\usepackage[usenames, dvipsnames]{color}

\newcommand\Ra{\mathrm{Ra}}

\newcommand\Nu{\mathrm{Nu}}

\renewcommand\L {\mathcal{L}}
\renewcommand{\citet}[1]{ref.~\cite{#1}}
\renewcommand{\Citet}[1]{Ref.~\cite{#1}}

\newcommand{\Citets}[1]{Refs.~\cite{#1}}

\renewcommand{\vec}[1]{\boldsymbol{#1}}

\newcommand{\eq}[1]{(\ref{#1})}
\newcommand{\eqs}[2]{(\ref{#1})~\&~(\ref{#2})}
\newcommand{\eqss}[2]{(\ref{#1})--(\ref{#2})}

\newcommand{\T}{{\cal T}}

\makeatletter
\let\Hy@backout\@gobble
\makeatother

\makeatletter
\newcommand*{\smallrel}[2][.8]{%
  \mathrel{\mathpalette{\smallrel@{#1}}{#2}}%
}
\newcommand*{\smallrel@}[3]{%
  \sbox0{$#2\vcenter{}$}%
  \dimen@=\ht0 %
  \raise\dimen@\hbox{%
    \scalebox{#1}{%
      \raise-\dimen@\hbox{$#2#3\m@th$}%
    }%
  }%
}
\makeatother

\begin{document}

\title{Marginally-Stable Thermal Equilibria of Rayleigh-B\'enard Convection}

\author{Liam O'Connor$^1$}
\author{Daniel Lecoanet$^{1, 2}$}
\author{Evan H. Anders$^2$}
\affiliation{%
$^1$Department of Engineering Sciences and Applied Mathematics, Northwestern University, Evanston, IL 60208 USA}
\affiliation{%
$^2$Center for Interdisciplinary Exploration and Research in Astrophysics, Northwestern University, Evanston, IL, 60201 USA}

\begin{abstract}
    Natural convection exhibits turbulent flows which are difficult or impossible to resolve in direct numerical simulations. In this work, we investigate a quasilinear form of the Rayleigh-B\'enard problem which describes the bulk one-dimensional properties of convection without resolving the turbulent dynamics. We represent perturbations away from the mean using a sum of marginally-stable eigenmodes.  By constraining the perturbation amplitudes, the marginal stability criterion allows us to evolve the background temperature profile under the influence of multiple eigenmodes representing flows at different length scales.
    We find the quasilinear system evolves to an equilibrium state where advective and diffusive fluxes sum to a constant.
    These marginally-stable thermal equilibria (MSTE) are exact solutions of the quasilinear equations.
    The mean MSTE temperature profiles have thinner boundary layers and larger Nusselt numbers than thermally-equilibrated 2D and 3D simulations of the full nonlinear equations.  
    MSTE solutions exhibit a classic boundary-layer scaling of the Nusselt number $\Nu$ with the Rayleigh number $\Ra$ of $\Nu \sim \Ra^{1/3}$.
    When an MSTE is used as initial conditions for a 2D simulation, we find that Nu quickly equilibrates without the burst of turbulence often induced by purely conductive initial conditions, but we also find that the kinetic energy is too large and viscously attenuates on a long viscous time scale.
\end{abstract}

\maketitle

\section{Introduction}
Rayleigh-B\'enard convection plays a foundational role in astrophysical and geophysical settings.
The resulting buoyancy-driven flows regulate heat transfer and generate large-scale vortices \cite{Couston}.
Turbulent convection, which is associated with large Rayleigh numbers $\Ra$, is difficult to simulate. 
State of the art simulations performed by \cite{Zhu_2018} have reached $\Ra \sim 10^{14}$ but estimates for the sun's convective zone and earth's interior are $\Ra \sim 10^{16}-10^{20}$ and $\Ra \sim 10^{20}-10^{30}$ respectively \cite{Ossendrijver,Gubbins_2001}. 
The scaling behavior of the Nusselt number $\Nu \sim \Ra^{\beta}$ in the asymptotic ultimate regime is of particular interest.
There is a substantial body of work pertaining to this specific topic with no general agreement between numerical simulations and asymptotic theories \cite{Malkus_1954, Howard_1966, Kraichnan, Spiegel, Castaing, Grossman, Ahlers}. 

Absent solid evidence from direct numerical simulation, other methods have been developed to try to infer large-$\Ra$ behavior or otherwise gain insight.
In the presence of other physical effects (e.g., rotation, magnetic fields), one can sometimes derive an asymptotically consistent set of reduced equations \cite{Julien2007, Julien2012}.
Reduced models are useful because they allow us to study the problem with less expensive computations.
Another approach relates to unstable exact coherent states (ECS) which are steady solutions to the full nonlinear problem \cite{Waleffe, Sondak, Wen, chini_cells}. 
Simulations and analyses performed by \cite{Yalniz, Cvitanovic} suggest that chaotic solution trajectories might intermittently resemble these ECS.
Should that be the case, it is crucial that we discover and classify such equilibria. 

Others have turned to studying quasilinear systems.
The quasilinear approximation starts with a decomposition of all variables into a background and perturbations about this background.
This approximation neglects the nonlinear interactions between perturbations which would otherwise modify the perturbations themselves \cite{marston2016}.
This renders the perturbation equations linear.
Although the quasilinear approximation greatly simplifies the problem, an additional condition must be imposed to determine the amplitude of the perturbations.

\Citets{herring63, marston2016} address this by initializing the perturbations to low amplitude and evolving them together with the background.
More recently, a different approach was taken in \citet{Beaume_2015}, which computes ECS in parallel shear flows by deriving and solving a quasilinear formulation of the Navier-Stokes equations via multi-scale asymptotic arguments. 
They assume the background velocity evolves on a slow time scale, and, to determine the perturbation amplitudes, they require marginal stability at each timestep.
% A similar strategy is employed by \citet{michel_chini_2019} to study acoustic streaming.
% In that work, an analytic expression for the first-order perturbation's amplitude is found by deriving a solvability condition.
This quasilinear approach, which relies on multi-scale arguments in conjunction with a marginal stability constraint, is generalized by \citet{Chini_ql}.
Studying two somewhat generic systems, researchers observe the interdependent evolution of a background state and its perturbations within a marginally-stable manifold.

In \citet{herring63}, Herring derives a quasilinear model for Boussinesq convection and provides supporting arguments for the model's physical validity.
He demonstrates that a single eigenmode is dominant only when $\Ra < 10^6$.
Herring uses numerical integration along with Fourier decomposition to approximate the transient solution to his model, allowing the system to saturate to a state of marginal stability.
Though efficient, this method is limited in its inability to model high-$\Ra$ convection due to the appearance of a second unstable mode at $\Ra \approx 10^6$.
How these modes interact to form thermal equilibria is also of interest.
In this paper we apply \citet{Chini_ql}'s generalized approach to \citet{herring63}'s reduced convection model by requiring marginal stability at the point of initialization.
When evolving the background state, we select the perturbation amplitudes such that marginal stability is maintained with each successive timestep.
This new method allows us to calculate marginally-stable thermal equilibria up to $\Ra = 10^9$, involving as many as five eigenmodes each with a unique horizontal wavenumber.

The paper is organized as follows: in Section~\ref{sec:model} we recall the underlying equations, and in Section~\ref{sec:evolution} we outline how we evolve the background temperature profile while maintaining marginal stability.
Section~\ref{sec:properties} pertains to the properties of the marginally-stable thermal equilibria, in particular how the Nusselt number and characteristic wavenumbers vary with the Rayleigh number.
Finally, we analyze the results of simulations initialized with marginally-stable thermal equilibria in Section~\ref{sec:sims}, and conclude in Section~\ref{sec:Discussion}.
 
\section{Model Setup}\label{sec:model}
We begin with the Boussinesq approximation for Rayleigh-Bénard Convection, nondimensionalized on the free-fall time scale. 
The domain $\mathcal{D}$ is 2-dimensional, rectangular, and horizontally periodic with cartesian spatial coordinates $0 \leq x < 4$ and $-1/2 < z < 1/2$.
We define the corresponding domain width $L_x = 4$ and height  $L_z = 1$.
The fluid of interest is constrained between two flat boundaries at $z = -1/2$ and $z = 1/2$ with fixed temperatures $1/2$ and $-1/2$ respectively. 
At both boundaries we specify impenetrable, no-slip conditions, such that the velocity $\vec{u} = u \hat{x} + w \hat{z} = \vec{0}$ at $z = \pm 1/2$, where $\hat{x}$ and $\hat{z}$ are the unit vectors in the $x$ and $z$ directions. 
The equations of motion are then given by
\begin{align}
    \nabla \cdot \vec{u} &= 0 \label{EQ:motion1}\\
    \frac{\partial \vec{u}}{\partial t} + \vec{u} \cdot \nabla \vec{u} &= - \nabla p + T \hat{z} + \mathcal{R} \nabla^2 \vec{u} \label{EQ:motion2}\\
    \frac{\partial T}{\partial t} + \vec{u} \cdot \nabla T &= \mathcal{P} \nabla^2 T \label{EQ:motion3}
\end{align}
where $p$ is pressure and $T$ is temperature. 
%For completeness, we specify a final boundary condition $p = p_0$ at $z = - 1/2$. 
A Boussinesq convection system of this form can be characterized by its dimensionless Rayleigh number $\Ra = \frac{g\alpha L^3_z \Delta T}{\nu \kappa}$ and Prandtl number $\Pr = \frac{\nu}{\kappa}$, where $g, \, \alpha,, \, \Delta T, \nu, \kappa$ are the gravitational acceleration, coefficient of thermal expansion, imposed temperature difference, kinematic viscosity, and thermal diffusivity respectively. 
In this paper, we fix $\Pr = 1$.
For convenience, we define
\begin{equation}
\mathcal{R} = \sqrt{\frac{\Pr}{\Ra}}, \qquad \mathcal{P} = \frac{1}{\sqrt{\Pr \Ra}}.
\end{equation}

To derive the quasilinear form, we posit that an arbitrary field $f$ can be represented as the sum of a mean profile (denoted by $\overline{f}$) and a perturbation function (denoted by $f'$)
\begin{align}
    \vec{u}(x, z, t) &= \vec{u'}(x, z, t) \label{EQ:reynolds_dc_u}\\
    &= u'(x, z, t)\hat{x} + w'(x, z, t)\hat{z} \\
    T(x, z, t) &= \overline{T}(z, t) + T'(x, z, t) \label{EQ:reynolds_dc_T}\\
    p(x, z, t) &= \overline{p}(z, t) +  p'(x, z, t) \label{EQ:reynolds_dc_p}
\end{align}
where the mean-velocity components vanish due to incompressibility and symmetry. Perturbations are defined to have no horizontal-average
\begin{equation}
    \langle f'(x, z, t) \rangle_x \equiv \int_{0}^{L_x} f'(x, z, t) dx = 0.
\end{equation}
%Assuming the existance of nontrivial solutions, we also fix the volume-average over the entire domain
%\begin{equation}
%  \langle T'^2 \rangle_{\mathcal{D}} = 1
%\end{equation}
%thereby alleviating any ambiguity in $A(t)$. 
Substituting \eq{EQ:reynolds_dc_T} into \eq{EQ:motion3} and taking the horizontal-average reduces the system to a simple initial value problem for $\overline{T}$
\begin{equation}
  \frac{\partial \overline{T}}{\partial t} + \frac{\partial}{\partial z} \langle w'T' \rangle_x = \mathcal{P} \frac{\partial^2 \overline{T}}{\partial z^2}, \label{EQ:T0_IVP}
\end{equation}
with associated boundary conditions $\overline{T}(-1/2, t) = 1/2$ and $\overline{T}(1/2, t) = -1/2$. It should be noted that we could obtain a similar equation for $u$ by breaking reflection symmetry about the midplane $z = 0$ and considering some nontrivial mean horizontal flow $\overline{u}(z, t)$. However we must have $\overline{w}(z, t) = 0$ due to incompressibility.

Substituting \eq{EQ:reynolds_dc_p} into \eq{EQ:motion2} and taking the horizontal average reveals that the mean pressure field $\overline{p}(z, t)$ must satisfy
\begin{equation}
    0 = -\frac{\partial \overline{p}}{\partial z} + \overline{T}. \label{EQ:p_bar}
\end{equation}

To solve \eq{EQ:T0_IVP} numerically, we need an expression for the perturbations so we can calculate the advective heat flux.
Here we will make the quasilinear approximation, dropping the $\vec{u}'\vec{\cdot}\vec{\nabla}\vec{u}'$ and $\vec{u}'\vec{\cdot}\vec{\nabla}T'$ terms from the evolution equations for the perturbations.
Substituting \eqss{EQ:reynolds_dc_u}{EQ:reynolds_dc_p} into \eqss{EQ:motion1}{EQ:motion3} followed by subtracting \eqs{EQ:T0_IVP}{EQ:p_bar} from the resulting temperature and $\hat{z}$ momentum equations respectively gives
\begin{align}
    \nabla \cdot \vec{u'} &= 0 \label{EQ:linear1}\\
    \frac{\partial\vec{u'}}{\partial t} &= - \nabla p' + T'\hat{z} + \mathcal{R} \nabla^2 \vec{u'} \label{EQ:linear2}\\
    \frac{\partial T'}{\partial t} + \frac{\partial \overline{T}}{\partial z} w' &= \mathcal{P} \nabla^2 T' \label{EQ:linear3}
\end{align}
with Dirichlet boundary conditions 
\begin{equation}
    T'|_{z = \pm \frac{1}{2}} = 0, \quad \vec{u'}|_{z = \pm \frac{1}{2}} = 0, \quad p'|_{z = \pm \frac{1}{2}} = 0.
\end{equation}
This is now a linear problem in $\vec{u}'$ and $T'$ which can be solved as an eigenvalue problem.

In his groundbreaking report \cite{Rayleigh_1916}, Lord Rayleigh observed that \eqss{EQ:linear1}{EQ:linear3} can be manipulated into a separable form with generalized solutions
\begin{align}
    w'(x, z, t) &= A\, \Re\left[W(z) \, e^{i(k_xx-st)}\right] \label{EQ:normal_modes1}\\ 
    u'(x, z, t) &= A\, \Re\left[U(z) \, e^{i(k_xx-st)}\right] \label{EQ:normal_modes2}\\ 
    T'(x, z, t) &= A\, \Re\left[\theta(z) \, e^{i(k_xx-st)}\right] \label{EQ:normal_modes3}\\ 
    p'(x, z, t) &= A\, \Re\left[P(z) \, e^{i(k_xx-st)}\right] \label{EQ:normal_modes4}
\end{align}
where $A$ is the (undetermined) mode amplitude, $s = \omega + i\sigma$ is the eigenvalue, and $k_x$ is constrained, by periodicity, to the countably infinite set (spectrum) of wavenumbers
\begin{align}
    k_x \in \left\{\frac{n\pi}{2} \, \big| \, n \in \mathbb{N}\right\}.
\end{align}
We normalize the eigenmodes to have
\begin{equation}
  \langle |\theta|^2 \rangle_{\mathcal{D}} = 1
\end{equation}
where $\langle \cdot \rangle_{\mathcal{D}}$ denotes the spatial mean over the entire domain.
Crucially, we emphasize that  \eqs{EQ:normal_modes1}{EQ:normal_modes3} are solutions to the linear problem.
In the quasilinear context, the background state and its eigenmodes vary in time.
By convention, our notation neglects this $t$-dependence in the eigenfunctions as they will always be computed from the autonomous linear system.

For each $k_x$, we can assess the stability of the perturbations by solving for the eigenvalue $s$, whose imaginary component $\sigma$ plays the role of an exponential growth rate. 
Positive eigenvalues indicate that the system is unstable to small disturbances whose Fourier decomposition includes a nontrivial component of wavenumber $k_x$.
Negative eigenvalues indicate stability. 
A complete linear stability analysis requires solution over the full spectrum of wavenumbers. 
The prototypical case is used to demonstrate that the critical Rayleigh number $\Ra_c = 1708$ when $\frac{\partial \overline{T}}{\partial z} = -1$.

To calculate the advective heat flux in equation~\ref{EQ:T0_IVP}, we can sum the individual advective terms $\langle w'T' \rangle_x$ of every marginally-stable mode.
In this way, the heat flux from the perturbations influence the evolution of $\overline{T}$.
But the evolution of $\overline{T}$ also influences the perturbations, as equation~\ref{EQ:linear3} depends on $\partial_z \overline{T}$.
Thus, the mean temperature and perturbation fields are coupled, as is the case in~\cite{Beaume_2015, Chini_ql}.

\section{Perturbation Evolution}\label{sec:evolution}
The linearized system \eqss{EQ:linear1}{EQ:linear3} does not constrain the amplitude of the eigenmodes, $A$.
However, the advective heat flux is proportional to $A^2$, so we need to specify the amplitude in order to solve equation~\ref{EQ:T0_IVP}.
To evolve $\overline{T}$, we assume the perturbations evolve on a much faster time scale than the mean temperature, as in \cite{michel_chini_2019}.
Stable modes $(\sigma < 0)$ decay away rapidly. 
Unstable modes $(\sigma > 0)$ will not persist on the slow time scale because the advective term $\langle w'T' \rangle_x$ tends to stabilize $\overline{T}$, thereby creating a negative feedback loop.
Only marginally-stable modes can be maintained on the slow time scale.
Therefore the amplitude $A$ must satisfy
\begin{equation}
    \max_{k_x} \{ \sigma \} = 0.
\end{equation}

For various $\Ra$ and fixed $\Pr = 1$, we seek marginally-stable thermal equilibria (MSTE) satisfying $\frac{\partial \overline{T}}{\partial t} = 0$ according to \eq{EQ:T0_IVP}. 
We employ the \texttt{Dedalus} pseudo-spectral python framework\footnotemark[1] \cite{Dedalus_2020} to solve the eigenvalue problem outlined in Section~\ref{sec:model} as well as the time evolution equation~\ref{EQ:T0_IVP}.
We represent each field with Chebyshev polynomials and use the 3/2 dealiasing rule to calculate the advective heat flux.
The necessary number of basis functions increases with $\Ra$ as the eigenfunctions include increasingly small-scale features (see Appendices \ref{sec:tables} \& \ref{sec:eigenfunctions}). 
We use the \texttt{Eigentools} package\footnotemark[2] \cite{Eigentools} to manipulate the eigenfunctions and calculate the advective heat flux $\langle w' T' \rangle_x$.

\footnotetext[1]{We employ \texttt{Dedalus} version v2.2006

\url{https://dedalus-project.org/}

\url{https://github.com/DedalusProject/dedalus/releases}
}

\footnotetext[2]{We employ \texttt{Eigentools} commit 

f457afc3193c32c16fdff94cdad962d07edea56b

\url{https://github.com/jsoishi/eigentools}}

To initialize our calculations, we construct a marginally-stable initial temperature profile $\overline{T}(z, t=0)$ whose equation is given in Appendix~\ref{sec:initial_profile}. 
Although this background temperature is marginally-stable, it is not in thermal equilibrium, so it will evolve in time.
We use \eq{EQ:T0_IVP} to evolve $\overline{T}(z, t_0)$ into a new marginally-stable profile $\overline{T}(z, t_0 + \Delta t)$.
We use a second-order, two-stage IMEX Runge-Kutta method.
The eigenfunctions and amplitudes are assumed to be constant with respect to $t$ over the timestep.
We symmetrize $\overline{T}$ by setting the coefficients of its even Chebyshev basis functions to zero at the start of each timestep.
This ensures $\overline{T}(z) = -\overline{T}(-z)$.
On every timestep, we calculate the marginally-stable eigenfunctions and their amplitude $A$.
It is essential that we pick the correct eigenfunction amplitude when calculating the advective term to maintain marginal stability.
We will now illustrate our method of finding the appropriate $A$ through an example.

\begin{figure}
    \includegraphics[width=3.375in]{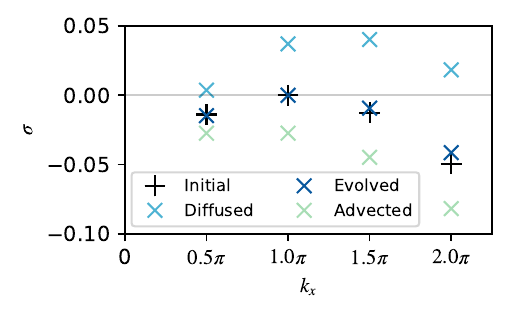}
    \caption{Eigenvalue spectra for $\Ra = 10^5$. The spectrum of an ``initial" marginally-stable mean temperature profile $\overline{T}(z, t_0)$ has a maximum eigenvalue of 0. 
    Given a small fixed timestep $\Delta t$, diffusion (equation~\ref{eqn:diffusion_only}) destabilizes the system, increasing $\sigma$. 
    Advection (equation~\ref{eqn:advection_only}) tends to stabilize the system, decreasing $\sigma$. 
    We find the eigenfunction amplitude $A^2$ such that the combination of diffusion and advection yields a new, ``evolved,'' marginally-stable mean temperature profile $\overline{T}(z, t_0 + \Delta t)$. In this case $A^2 \approx 1$ due to approximately equal magnitudes of the diffused and advected eigenvalues.}
    \label{fig:iteration_spectra} 
\end{figure}

Consider a marginally-stable temperature profile $\overline{T}(z, t_0)$.
By definition, its maximum growth rate is 0.
Diffusing $\overline{T}(z, t_0)$ tends to increase its growth rates while ignoring the diffusive term and evolving according to advection tends to decrease its growth rates (see Figure~\ref{fig:iteration_spectra}).
The amplitude $A$ must be selected such that these two influences are equal and opposite.
We can measure the effects of diffusion and advection on the maximum growth rate by solving two new initial value problems
\begin{align}\label{eqn:diffusion_only}
    &\frac{\partial \overline{T}_{\rm{diff}}}{\partial t} = \mathcal{P} \frac{\partial^2 \overline{T}_{\rm{diff}}}{\partial z^2} \\
    &\frac{\partial \overline{T}_{\rm{adv}}}{\partial t} + 2\frac{\partial}{\partial z} \Re\left[ W \theta^* \right] = 0 \label{eqn:advection_only}
\end{align}
where $\overline{T}_{\rm{diff}}$ and $\overline{T}_{\rm{adv}}$ denote the diffused and advected temperature profiles respectively. 
In equation~\ref{eqn:advection_only} we have set the amplitude of the eigenfunctions $A=1$ so we can determine how advection changes the temperature profiles.
The factor of 2 arises due to the horizontal averaging of the advective term in \eq{EQ:T0_IVP}.

We solve equations \eqs{eqn:diffusion_only}{eqn:advection_only} from $t=t_0$ to $t=t_0+\Delta t$, initializing each temperature profile with $\overline{T}(z, t_0)$.
Suppose $k_0$ is the wavenumber of the marginally-stable mode.
We then calculate the $k_x = k_0$ mode's growth rates $\sigma_{\rm{diff}}$ and $\sigma_{\rm{adv}}$ for $\overline{T}_{\rm{diff}}$ and $\overline{T}_{\rm{adv}}$.
A good guess for $A^2$ is
\begin{equation}
    A^2 \approx \Lambda^2_0 \equiv -\frac{\sigma_{\rm{diff}}}{\sigma_{\rm{adv}}}. \label{EQ:amp_approx}
\end{equation}
This estimate follows from first-order perturbation theory.
Note that $\sigma_{\rm{adv}} < 0$ and $\sigma_{\rm{diff}} > 0$, so $\Lambda_0^2>0$.

Using the approximate amplitude $\Lambda_0^2$, the background temperature $\overline{T}(z, t_0)$ is then evolved to $t=t_0+\Delta t$ according to \eq{EQ:T0_IVP} and another eigenvalue solve is performed. 
This background temperature is typically close to, but not exactly, marginally-stable.
In fact, in the limit $\Delta t \rightarrow 0$, the estimate $\Lambda_0^2$ approaches the amplitude $A^2$ necessary to keep the background temperature marginally-stable \cite{Chini_ql}.
We use Newton's method to find an amplitude $A^2$ such that the maximum growth rate is zero to within $10^{-9}$.
We observe that marginally-stable modes do not oscillate in time, i.e.~$\sigma = 0$ implies $\omega = 0$.
This numerical result is consistent with the conventional notion of exchange of stabilities \cite{drazin_reid_2004}.
Crucially, we do not assume the $k_x$ of the marginally-stable mode is fixed.
In Section~\ref{sec:multiple_modes} we specify procedures for the treatment of multiple simultaneously marginal modes.

\subsection{Treatment of Multiple Marginally-Stable Modes} \label{sec:multiple_modes}
In most cases, we encounter eigenvalue spectra with multiple marginal modes.
To accommodate this we generalize the advective term in \eq{EQ:T0_IVP} to accommodate $N$ simultaneously marginal modes
\begin{equation}
    \langle w' T' \rangle_x = \sum_{n = 1}^{N} 2 A_n^2  \Re\left[ W_n \theta_n^* \right] 
\end{equation}
where $W_n$ and $\theta_n$ are the marginally-stable eigenmodes associated with different values of $k_x$. 
The factor of 2 is again due to horizontal averaging.
There are now $N$ modes, each with their own amplitude to solve for and eigenvalue to keep marginally-stable. 
Given a small fixed time step $\Delta t$, let $\vec{A^2}$ be the amplitude vector and $\vec{\sigma}(\vec{A^2})$ be the eigenvalues of the $N$ modes.
We are searching for $\vec{\tilde{A}^2}$ that satisfies $\vec{\sigma}(\vec{\tilde{A}^2})=0$.
We approximate the amplitudes with $\vec{\Lambda^2_0}$ by generalizing \eq{EQ:amp_approx}:
\begin{equation}
    \vec{\tilde{A}^2} \approx \vec{\Lambda^2_0} = -\Sigma_{\rm{adv}}^{-1} \vec{\sigma_{\rm{diff}}}.
    \label{EQ:AN_approx}
\end{equation}
Here $\vec{\sigma_{\rm{diff}}} = \vec{\sigma} (\vec{0})$ refers to the eigenvalues of the background temperature after evolving under only diffusion (equation \ref{eqn:diffusion_only}) for a time interval $\Delta t$.
We account for the influence of $N$ advection terms on $N$ eigenvalues (one for each mode) by constructing an eigenvalue matrix $\Sigma_{\rm{adv}}$.
To calculate the element of $\Sigma_{\rm{adv}}$ at row $i$ and column $j$, we evolve the background temperature under only advection by mode $j$ for a time interval $\Delta t$.
Then the $i,j$ component of $\Sigma_{\rm{adv}}$ is given by the eigenvalue of the $i$th mode.

Our estimate $\vec{\Lambda^2_0}$ does not typically give a background temperature for which all $N$ modes are exactly marginally-stable.
We refine our vector of amplitudes via Newton's method.
This requires calculating the Jacobian matrix 
\begin{equation}
    J = \begin{bmatrix}
        \nabla \sigma_1 (A_1, A_2, ..., A_N) \\
        \nabla \sigma_2 (A_1, A_2, ..., A_N) \\
        \vdots \\
        \nabla \sigma_N (A_1, A_2, ..., A_N) 
    \end{bmatrix}.
\end{equation}
We approximate $J$ via first-order finite differences, which requires an additional $N^2$ sparse eigenvalue solves.
%A once-refined estimate $\vec{A^2_1}$ is then given by
%% The $k$th row of $J_0$ is then given by the gradient of the $k$th mode's eigenvalue $\nabla \sigma_k$.
%\begin{equation}
%    \vec{\tilde{A}^2} \approx \vec{A^2_1} = -J^{-1} \vec{\sigma}(\vec{A^2_0}).
%\end{equation}
%We then use each subsequent guess to adjust the Jacobian via Broyden's method for root-finding in multi-dimensional functions \cite{Broyden}.
We iterate Newton's method until all marginally-stable modes have eigenvalues within $10^{-9}$ of zero.
We find $\vec{\sigma}(\vec{A^2})$ does indeed have a unique root provided the time step is not too large and there are no numerical instabilities, as outlined in Appendix~\ref{sec:timestep}.
Presumably this is due to the coupling of $\langle w'T' \rangle_x$ with $\overline{T}$.
Over the course of a large time step, $\overline{T}$ evolves according to \eq{EQ:T0_IVP} and eventually the original eigenfunctions cease to provide a stabilizing influence.
This limits the timestep size we can take.

Difficulty arises when transitioning between different numbers of marginal modes.
We facilitate these transitions by defining an adjustable candidate tolerance $\varepsilon_{\rm{cand}} \in [10^{-6}, 10^{-8}]$.
A mode which meets the candidate tolerance can be included in the subsequent timestep as a candidate marginal mode.
Candidate modes are rejected when the root-finding algorithm converges on a negative amplitude $A^2 < 0$.
Otherwise the mode becomes marginally-stable.
In a similar manner, a marginally-stable mode need not remain marginally-stable.
If its amplitude converges to some $A^2 < 0$ as before, then that mode is discarded and the timestep is repeated.
\section{Properties of Thermally Equilibrated States}\label{sec:properties}
We evolve $\overline{T}$ as described above until $\max|\partial_{t}\overline{T}| < 10^{-5}$.
In this marginally-stable thermal equilibrium, $\overline{T}$ does not evolve in time, and the perturbations also do not evolve in time, as they are marginally-stable.
Thus, these configurations are exact solutions to the quasilinear system (equations~\ref{EQ:T0_IVP}--\ref{EQ:linear3}).
They differ from the usual ECS in that ECS are fixed points of the full nonlinear problem \eqss{EQ:motion1}{EQ:motion3}. 
Such definitions are not mutually exclusive, but in general we can assume that MSTE and ECS are not steady with respect to their counterparts' equations.
We compute symmetric MSTE for $\Ra$ in the range $10^5 - 10^9$.

\begin{figure}
    \centering
    \includegraphics[width=3.375in]{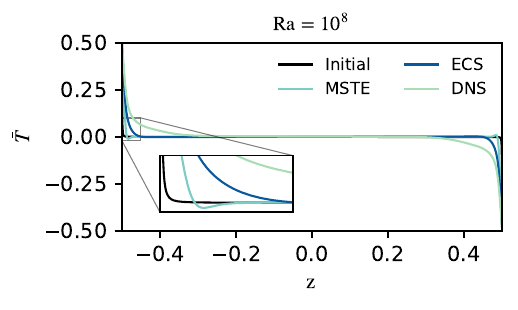}
    \caption{Mean temperature profiles $\overline{T}$ for $\Ra = 10^8$. 
    The initial profile is given by \eq{EQ:T0}. 
    We evolve this background temperature profile according to Section~\ref{sec:evolution} until we reach a marginally-stable thermal equilibrium (MSTE).
    ECS refers to a steady nonlinear solution for Boussinesq convection, given by \citet{Wen}.
    The DNS curve is obtained from a 2D nonlinear simulation of \eqss{EQ:motion1}{EQ:motion3} with \texttt{Dedalus}.
    DNS temperature data are horizontally- and time-averaged.
    The initial profile has the narrowest boundary layer, while the DNS profile has the widest boundary layer.
    The equilibrated MSTE and ECS curves lie between the initial and DNS curves, with the ECS having a more diffuse boundary layer than that of the MSTE.
    The MSTE profile exhibits prominent dips, nested alongside the boundary regions. 
    }
%    The source of this feature is not well understood, but similar temperature gradient reversal regions were found by \cite{chini_cells} along the midlines of 2D convective cellular solutions at $\Ra \sim 10^6$.}
    \label{fig:T0_profiles}
\end{figure}

Figure~\ref{fig:T0_profiles} gives temperature profiles for $\Ra = 10^8$, where the initial profile is outlined in Appendix~\ref{sec:initial_profile}. 
We run a direct numerical simulation (DNS) of \eqss{EQ:motion1}{EQ:motion3} with \texttt{Dedalus} and plot the horizontal- and time- averaged temperature profile. 
The temperature boundary layers in the DNS are wider than in the MSTE, which is in turn wider than the boundary layer of the initial temperature profile.
By computing the maximum growth rates of the linearized system \eqss{EQ:linear1}{EQ:linear3}, we found that both the DNS and the ECS profiles are linearly unstable when used as background states.
Every marginally-stable initial profile we experiment with ($\tanh$, $\rm{erf}$) has thinner boundary layers than the DNS and MSTE profiles.
When comparing our quasilinear equilibrium (MSTE) with \citet{Wen}'s nonlinear equilibrium (ECS), we find that the ECS profile is significantly more diffuse than that of the MSTE.
In conjunction, these two facts suggest that MSTE maximize boundary layer thickness while subject to the marginal stability constraint.

\begin{figure*}
    \centering
    \begin{tabular}{@{}c@{}}
        \includegraphics[width=3.375in]{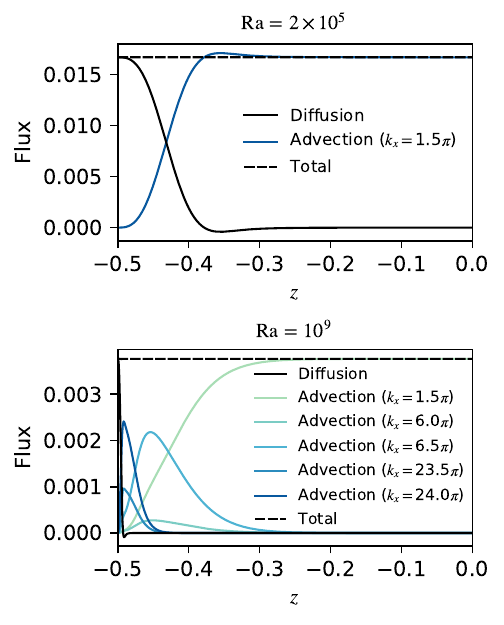}
    \end{tabular}
    \begin{tabular}{@{}c@{}}
        \includegraphics[width=3.375in]{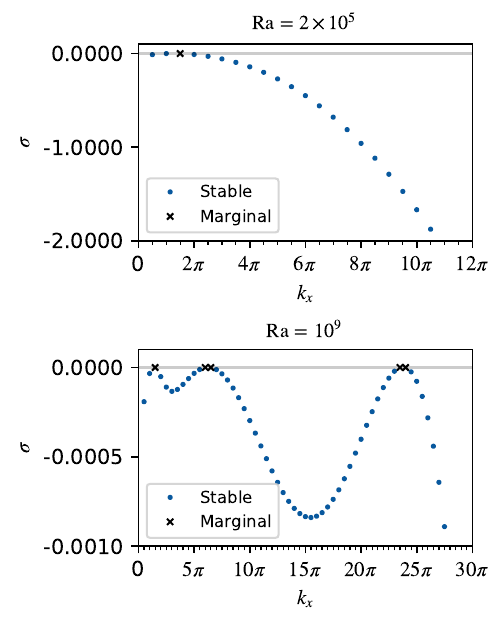}
    \end{tabular}
    \caption{Heat fluxes (left) and eigenvalue spectra (right) of equilibrated states with $\Ra = 2 \times 10^5$ (top) and $\Ra = 10^9$ (bottom). 
    The heat flux profiles are symmetric about $z=0$, so we only plot for $z<0$.
    We plot the advection profiles for marginally-stable modes. 
    At low $\Ra$, a single mode with $k_x = 1.5\pi$ is sufficient to oppose boundary layer diffusion and facilitate heat flux throughout the bulk of the domain. 
    At large $\Ra$, high-wavenumber modes contribute pronounced small-scale advection profiles which tightly hug the thin boundary layers. 
    A combination of progressively wider advection profiles is necessary to transition to bulk advection provided by the $k_x = 1.5\pi$ mode. 
    This bulk mode mimics the large-scale convective cells which are characteristic of Rayleigh–Bénard convection, both in qualitative structure and number of cells in the domain.
    The eigenfunctions which compose the advective profiles are illustrated in Appendix~\ref{sec:eigenfunctions}.}
    \label{fig:flux}
\end{figure*}

The most resilient and unexpected feature of MSTE temperature profiles are the pronounced dips adjacent to the boundary layers. 
These dips appear in every solution, regardless of $\Ra$. 
Physically, they correspond to thin layers in which the mean temperature gradient reverses, contradicting an important hypothesis of \cite{Malkus_1954}. 
This counter-diffusion, which opposes overall heat transfer, is overcome by the coinciding advective flux, shown in Figure~\ref{fig:flux}. 
\Citet{herring63} observes this when studying the same system.
Similar temperature gradient reversals are reported by \cite{chini_cells} along the midlines of 2D convective cellular solutions at $\Ra \sim 10^6$. 
In that case, the reversals are due to nonlinear advection, which is not present in our quasilinear model.
Neither the present authors, nor Herring were able to provide a rigorous mathematical explanation for this behavior.

In Figure~\ref{fig:flux}, we give heat flux profiles and eigenvalue spectra for two cases: $\Ra = 2 \times 10^5$ (top) and $\Ra = 10^9$ (bottom). 
Appendix~\ref{sec:eigenfunctions} shows representative vertical velocity and temperature perturbation eigenfunctions.
For $\Ra = 2 \times 10^5$, there is a single marginal mode at $k_x = 1.5\pi$ whose advective flux occupies the bulk of the domain. 
These states have wide boundary layers, with diffusive fluxes which gradually subside as advection becomes the dominant flux component. 
Transitional regions occur over a smaller length scale for $\Ra = 10^9$ where the shift from diffusion to advection is sharp.
At $\Ra=10^9$ we find five marginally-stable modes are necessary to reach an MSTE.
Thin advection profiles, belonging to high-wavenumber modes with $k_x=23.5\pi, \, 24\pi$, hug the boundary layer. 
Closer to the bulk of the domain, we see wider advection profiles corresponding to modes in a second pair of marginal modes $k_x = 6\pi, \, 6.5\pi$.
Adjacent pairs of marginal modes whose wavenumbers differ by $0.5\pi$ are common among MSTE.
The isolated $k_x = 1.5\pi$ mode forms the same arrangement and number of large-scale convective cells observed in DNS, again occupying the bulk of the domain.
The pairs of modes $k_x = 6\pi, \, 6.5\pi$ and $k_x=23.5\pi, \, 24\pi$ are each associated with a single maximum in our plots of growth rate $\sigma$ as a function of $k_x$ (lower right panel of Figure~\ref{fig:flux}).
If we allowed wavenumbers to vary continuously, there would be an unstable mode between these pairs of wavenumbers.
However, since we have fixed the horizontal size of our domain, we are left with pairs of discrete marginal modes.

\begin{figure}
    \centering
    \includegraphics[width=3.375in]{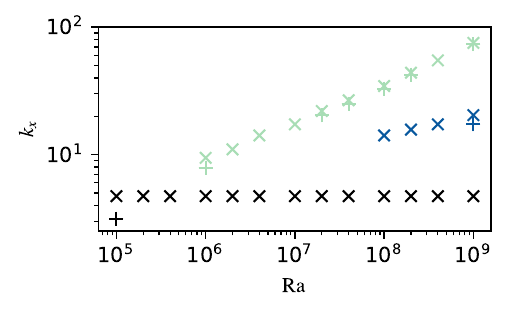}
    \caption{Wavenumbers of marginally-stable modes in thermally equilibrated states.
    Marginal modes often appear in adjacent pairs, which we denote with a common color. 
    Within each of these pairs, the largest and smallest $k_x$ are denoted with an x and a + respectively.
    For example, the spectrum corresponding to $\Ra = 10^5$ has adjacent marginal wavenumbers $k_x = \pi, \, 1.5\pi$. 
    The $\Ra = 10^9$ spectrum, shown in the lower right corner of Figure~\ref{fig:flux}, has three groups of maxima, with a single marginal mode in the first group ($k_x = 1.5\pi$), two adjacent marginal modes in the second group ($k_x = 6\pi, \, 6.5\pi$), and two adjacent marginal modes in the third group ($k_x = 23.5\pi, \, 24\pi$). 
    The largest wavenumbers of the green and blue branches depend on  $\Ra$ according to power laws.
    For the green branch, $\max\{k_x\} \propto \Ra^{0.300}$ with $R^2 = 0.998$ while for the blue branch, $k_x \propto \Ra^{0.155}$ with $R^2 = 0.988$.
    }
    \label{fig:kx_marginals}
\end{figure}

MSTE for large $\Ra$ tend to have a diverse combination of marginal modes.
In every case, the $k_x = 1.5\pi$ mode is included. 
In Figure~\ref{fig:kx_marginals} we give the wavenumbers $k_x$ of marginal modes. 
Adjacent pairs of marginal modes are common but not ubiquitous.
Like $k_x = 6\pi, \, 6.5\pi$ and $k_x=23.5\pi, \, 24\pi$ for the $\Ra=10^9$ MSTE, these are due to the discretization of wavenumbers from our domain of width 4.
We think of the pairs of modes as acting together as part of a single maximum of the growth rate as a function of the wavenumber.
When wavenumbers are adjacent, we plot them in the same color and denote the larger mode with an x and the smaller with a +.
For $\Ra \geq 10^6$, a second branch of marginal modes is shown in light green. 
Least-squares regression gives $\max\{k_x\} \propto \Ra^{0.300}$ with $R^2 = 0.998$ for this maximum branch.
This relationship does not appear to be asymptotic. 
If we consider only those MSTE whose $\Ra \geq 10^8$, we observe $\max\{k_x\} \sim \Ra^{1/3}$ in agreement with classical scaling arguments. 
For large $\Ra$, the advective fluxes of this maximum branch compensate for the strongly peaked diffusive flux in the thin boundary layers.
At $\Ra \geq 10^8$, a third branch appears (shown in blue), splitting the widening gap between the other two. 
For these points regression gives $k_x \propto \Ra^{0.155}$ with $R^2 = 0.988$.
The blue branch is associated with moderately wide advection profiles, filling a niche in the total flux by uniting the thin profiles of the maximum branch with those of the bulk-domain-oriented minimum branch.

\begin{figure}
    \centering
    \includegraphics[width=3.375in]{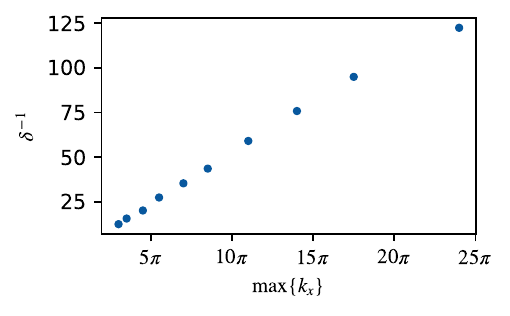}
    \caption{For $\Ra \geq 10^6$, the maximum marginally-stable wavenumber (corresponding to the green x markers in Figure~\ref{fig:kx_marginals}) are inversely related to the boundary layer height $\delta$. 
    $(\max \{ k_x \})^{-1}$ gives a minimum $x$ length scale for the perturbations, and consequently, the advection. 
    For large $\Ra$, the boundary layers admit small scale features, requiring more vertical basis functions (higher resolution, see Appendix~\ref{sec:tables} for specifications).
    The boundary layer height gives an estimate of the minimum vertical length scale in the problem.
    This suggests that the minimum horizontal and vertical length scales are proportional to each other over a wide range of $\Ra$.}
    \label{fig:del_inv}
\end{figure}

The largest marginal wavenumber $\max \{ k_x \}$ (represented by the light green branch in Figure~\ref{fig:kx_marginals}) serves as an inverse minimum length scale in the $x$ direction.
The finest vertical structures in $\overline{T}(z)$ appear near the boundaries, requiring more basis functions (resolution) at large $\Ra$.
Naturally, this provides a complementary minimum length scale for $z$.
We define the boundary layer height $\delta$ as the distance from the boundary where $\partial_z \overline{T}$ equals zero, i.e.,
\begin{align}\label{eqn:delta}
    \left.\frac{\partial \overline{T}}{\partial z}\right|_{z=-\frac{1}{2}+\delta} = 0, \qquad 0 < \delta < 1/2.
\end{align}
This height corresponds to the local extrema of the MSTE temperature profile, e.g., in Figure~\ref{fig:T0_profiles}.
In Figure~\ref{fig:del_inv}, we show $\max \{ k_x\}$ is proportional to $\delta^{-1}$ over our range of $\Ra$.
Least-squares regression gives $\delta^{-1} = 1.71 \max \{ k_x \} - 2.13$ with $R^2 = 0.996$.
% We can assume from this length scale agreement that the mean-squared $x$ and $z$ components of the temperature and velocity gradients are proportional
% \begin{align}
%     \Big\langle \frac{\partial T}{\partial x} \cdot \frac{\partial T}{\partial x} \Big\rangle_{\mathcal{D}} &\propto \Big\langle \frac{\partial T}{\partial z} \cdot \frac{\partial T}{\partial z} \Big\rangle_{\mathcal{D}} \nonumber   \\
%     \Big\langle \frac{\partial \vec{u}}{\partial x} \cdot \frac{\partial \vec{u}}{\partial x} \Big\rangle_{\mathcal{D}} &\propto \Big\langle \frac{\partial \vec{u}}{\partial z} \cdot \frac{\partial \vec{u}}{\partial z} \Big\rangle_{\mathcal{D}}.
% \end{align}
% This is consistent with the mean squared gradient assumptions of \cite{Malkus_1954}.
%The quasilinear model cannot describe arbitrarily small scale structures. This is a consequence of us collapsing the problem into a single dimension via mode-discretization.

\begin{figure}
    \centering
    \includegraphics[width=3.375in]{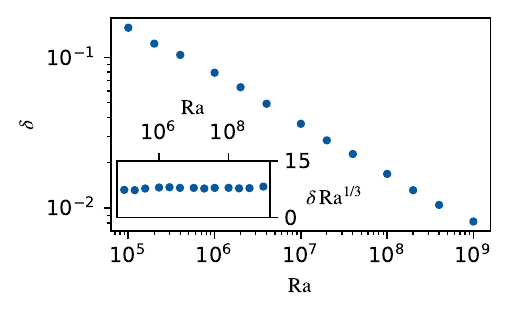}
    \caption{Boundary layer height $\delta$ of MSTE. 
    We define the boundary layer height based off the location where $\frac{\partial \overline{T}}{\partial z} = 0$ (see equation~\ref{eqn:delta}). 
    Plotting on a log-log scale, we find that $\delta$ and $\Ra$ obey a power-law relationship. We also demonstrate that $\Ra^{1/3}\delta$ is approximately constant with respect to $\Ra$ which is consistent with \cite{Malkus_1954}}
    \label{fig:bl_ra}
\end{figure}

We find the MSTE can be characterized by their boundary layer height.
In Figure~\ref{fig:bl_ra}, we illustrate the scaling behavior of the boundary layer height $\delta \sim \Ra^{-1/3}$. 
This is consistent with Malkus' classical marginal stability theory, a scaling argument which perceives the boundary regions as subdomains which are themselves marginally-stable \cite{Malkus_1954}.

\begin{figure}
    \centering
    \includegraphics[width=3.375in]{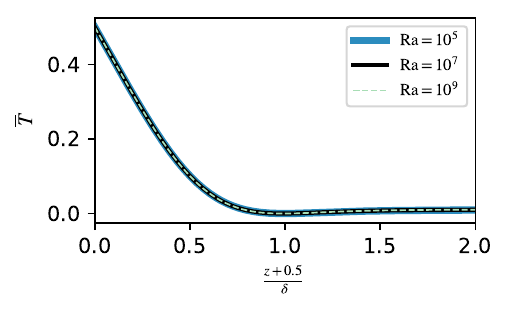}
    \caption{Temperature boundary layer geometries of various MSTE. 
    Rescaling according to the previously defined boundary layer height $\delta$ reveals self-similarity in the temperature boundary layers despite large variation in $\Ra$. 
    The rescaled curves remain approximately self-similar for $z < 0$.
    }
    \label{fig:b0_delta}
\end{figure}

The temperature boundary layers of MSTE exhibit self-similarity despite great variation in $\Ra$. 
We illustrate this in Figure~\ref{fig:b0_delta} where the mean temperature $\overline{T}$ is plotted along a rescaled $z$-coordinate $\frac{z + 1/2}{\delta}$.
It is also noteworthy that we do not encounter any appreciable variation in the boundary layer geometry as we extend Herring's findings beyond $\Ra = 10^6$.
The rescaled boundary layer geometry appears intrinsic to the MSTE, possibly minimizing diffusion as previously hypothesized.
This again highlights the importance of the boundary layer height's dependence on $\Ra$.

\begin{figure*}
    \centering
    \includegraphics[width=6.75in]{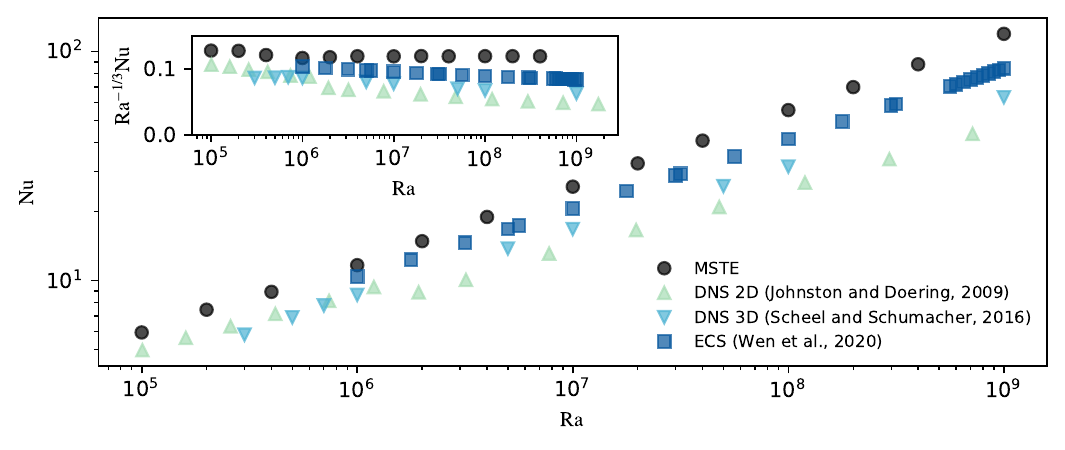}
    \caption{Nusselt numbers are shown for MSTE, aspect-ratio-optimized ``steady rolls'' ECS \cite{Wen}, as well as statistically-steady 2D and 3D DNS \cite{Johnston, Scheel_2016}.
    All datasets obey power-law relationships, with the MSTE and ECS scaling like $\Nu \sim\Ra^{1/3}$. 
    MSTE have greater $\Nu$ than the ECS, which in turn, have greater $\Nu$ than the DNS. 
    This can be explained by the contrasting boundary layer geometries shown in Figure~\ref{fig:T0_profiles}.}%
    \label{fig:nu_vs_ra}%
\end{figure*}

The Nusselt number, which measures convective performance is given by
\begin{equation}
    \Nu = \frac{\langle \langle w'T' \rangle_x - \mathcal{P}\frac{\partial \overline{T}}{\partial z} \rangle_z}{\langle- \mathcal{P}\frac{\partial \overline{T}}{\partial z} \rangle_z}.
\end{equation}
There is no general consensus surrounding the scaling behavior of $\Nu$ for high $\Ra$ systems, which are of particular importance in astrophysical and geophysical systems. In Figure~\ref{fig:nu_vs_ra} we report $\Nu$ for MSTE, ``steady rolls" ECS \cite{Wen}, and DNS \cite{Scheel_2016, Johnston}. 
We find that MSTE satisfy $\Nu \sim\Ra^{1/3}$, consistent with our finding that the boundary layer height scales like $\delta \sim \Ra^{1/3}$.
Herring was able to demonstrate this for $\Ra \leq 10^6$.
Our results confirm this trend in MSTE up to $\Ra = 10^9$.
The Nusselt numbers of the ECS are somewhat lower and the DNS Nusselt numbers are yet lower still.
In both cases, the $\Ra$ dependence appear slightly more shallow than for the MSTE.
\Citet{Wen} hypothesized that the $\Nu$ of all ECS which admit classical Malkus scaling must always exceed the $\Nu$ of turbulent convection.
If we generalize this notion to include quasilinear equilibria, our findings agree; MSTE have larger $\Nu$ than 2D and 3D DNS.
This might be due to the chaotic transitions among the unstable periodic orbits outlined by \cite{Yalniz, Cvitanovic} inhibiting heat flux. 
We might also anticipate the existence of similar equilibria with smaller $\Nu$, occupying complementary nodes in the Markov chain whose behavior agrees with DNS. 

\section{Simulations with Thermally Equilibrated Initial Conditions}\label{sec:sims}
This investigation is partially motivated by the prospect of decreasing DNS runtimes by employing MSTE as initial conditions. 
One common choice of initial conditions for DNS of equations \eqss{EQ:motion1}{EQ:motion3} are
\begin{align}
    T(x, z)\big|_{t=0} &= 0.5 - z + \mathcal{N} \nonumber \\
    \vec{u}(x, z)\big|_{t=0} &= \vec{0} \label{EQ:linear_ic}
\end{align}
where $\mathcal{N}$ is low-amplitude random noise concentrated in the bulk of the domain.
Here we instead initialize using the MSTE,
\begin{align}
    T(x, z)\big|_{t=0} &= \overline{T}(z) + \sum_{n=1}^N  A_n \Re \Big[ \theta_n(z) e^{ik_{x_n}x} \Big] + \mathcal{N} \nonumber \\
    \vec{u}(x, z)\big|_{t=0} &= \sum_{n=1}^N A_n \Re \Big[\Big( U_n (z) \hat{x} + W_n(z) \hat{z} \Big) e^{ik_{x_n}x} \Big] \label{EQ:mste_ic}
\end{align}
where $\theta_n(z), \, U_n(z), \, W_n(z); \; A_n; $ and $k_{x_n}$ refer to the complex eigenfunctions, amplitude, and wavenumber of the $n$th marginal mode respectively. 
Note that although the MSTE is an equilibrium of the quasilinear equations, it is not an equilibrium of the full nonlinear equations.
Accordingly the simulation state would evolve on initialization absent a random noise term.
Here we include noise as a source of asymmetries. 
We also perform a simulation with MSTE absent any initial velocity ($\vec{u}(x, z)|_{t=0} = \vec{0}$). 
This state is not in equilibrium but we refer to it as ``MSTE No Flow'' for clarity.

\begin{figure}
    \centering
    \includegraphics[width=3.375in]{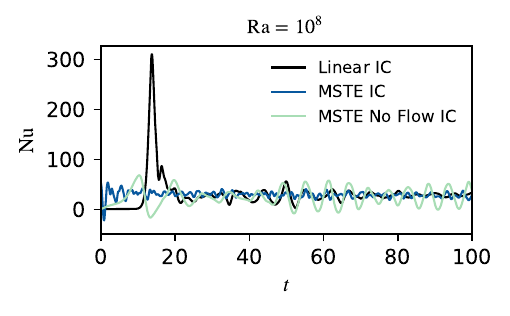}
    \caption{Nusselt numbers of simulations performed at $\Ra = 10^8$, initialized with the conductive equilibrium plus thermal noise (black), the MSTE (blue), and the MSTE absent initial flow (green). 
    The MSTE simulation does not undergo a convective-transient period because the characteristic large-scale convective cell structure exists on initialization.
    The MSTE no flow simulation exhibits quasi-periodic oscillations in $\Nu$ which appear to persist for long times.
    The oscillations are great enough in amplitude that $\Nu$ is quasi-periodically negative, indicating brief periods of average heat transfer reversal.}
    \label{fig:nu_sim}
\end{figure}

Simulations initialized with the conductive equilibrium plus low-amplitude thermal noise have a large peak in $\Nu$ early on in their evolution (Figure~\ref{fig:nu_sim}).
This is due to a burst of turbulence which occurs when the convective motions first become nonlinear.
The MSTE no flow simulation undergoes a similar transient period, albeit less pronounced as illustrated by the smaller peak in the green curve.
A simulation initialized with the MSTE, however, does not exhibit this transient burst of turbulence, as the large-scale anatomy of convective cells exists on initialization. 
Simulation of this transitional period is prohibitive \cite{Anders_AE}. 
For high $\Ra$ experiments, researchers often ``bootstrap" data by initializing simulations with the results of similar $\Ra$ runs \cite{Verzicco, Johnston}. 
MSTE can be perceived as a set of initial conditions, designed for avoiding the simulation of transient high Reynolds number flows.
MSTE absent velocity effectively achieve the same goal. 

MSTE are laminar, lacking the small-scale structures associated with moderate- to high-$\Ra$ experiments. 
This is an apparent consequence of the quasilinear assumptions. 
If we perceive MSTE as background states, DNS suggest that plumes, vortex sheets, and other unstable turbulent features inhibit total heat transfer. 
This perspective agrees with conventional models of transitions to turbulent flows, such as Boussinesq's turbulent-viscosity hypothesis \cite{pope_2000}. 
The emergence of small-scale velocity structures is indicative of nonlinear energy transfer to small scales where energy is lost due to viscosity. 
Buoyancy driven flows are therefore impeded and the advection in the bulk of the domain decreases in magnitude \cite{drazin_reid_2004, pope_2000}. 
We could also attribute the diffuse DNS temperature profile in Figure~\ref{fig:T0_profiles} with unsteady boundary-layer penetration and mixing that MSTE do not exhibit.

\begin{figure}
    \centering
    \includegraphics[width=3.375in]{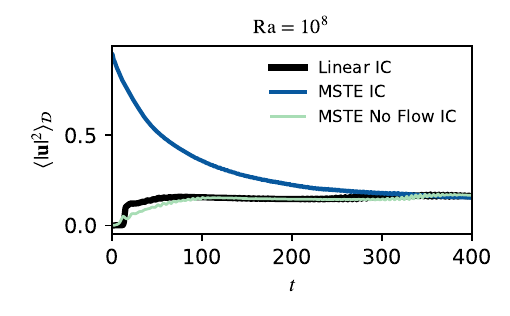}
    \caption{Average kinetic energies are reported for the same simulations illustrated in Figure~\ref{fig:nu_sim} ($\Ra = 10^8$). 
    The eigenfunctions belonging to MSTE have significantly more kinetic energy than the statistically-steady state. 
    Kinetic equilibrium is achieved on a long viscous time scale.}
    \label{fig:ke_sim}
\end{figure}

We also find that the average kinetic energy of the simulation initialized with MSTE is significantly larger than those found in either of the other cases, as shown Figure~\ref{fig:ke_sim}. 
This is because MSTE contain strong large-scale high-velocity flows, which decay on a long viscous time scale $t_{\nu}$. 
These powerful advective flows balance the heat flux in the bulk of the domain with the strong diffusion of thin MSTE temperature boundary layers.
Fluid in the other cases is initialized at rest. 
Two viscous time scales can be derived for this problem: one for the boundary layer 
\begin{align}
    t_{\nu,\delta} &= \frac{\delta^2}{\mathcal{P}} = 2.86
    \intertext{and another for the entire domain}
    t_{\nu,L} &= \frac{L^2_z}{\mathcal{P}} = 10^4.
\end{align}
Empirically, we find $t_{\nu} \approx 200$ suggesting the true kinetic energy decay time scale is based on a length scale $\ell \approx 0.14$ which is longer than the boundary layer height $\delta$ but shorter than the domain height $L_z$.
Consequently, MSTE initial conditions do not reduce the simulation time required to achieve a statistically-steady state---rather they increase it considerably!
This is alleviated by using no-flow MSTE initial conditions.
The MSTE temperature state is a promising candidate initial condition for high-Rayleigh number convection simulations.

\section{Discussion}\label{sec:Discussion}
In this paper we describe a new way to study Rayleigh–Bénard convection. 
We compute marginally-stable thermal equilibria (MSTE), which are equilibria of the quasilinear equations.
To compute MSTE, we construct a marginally-stable mean temperature profile and evolve it according to the advective flux of its marginally-stable eigenfunctions, and its own diffusion. 
We assume that at least some modes are always in a marginally-stable configuration and all other modes are stable. 
The marginal stability constraint then fixes the ratio between advection and diffusion (eigenfunction amplitude $A^2$). 
We use standard root-finding algorithms to solve for the appropriate $A^2$ at each timestep until the  combination of diffusive and advective flux sum to a constant.
The MSTE calculation is a one-dimensional problem, combining eigenvalue solves, and the time evolution of the one-dimensional mean temperature profile.
Thus, MSTE can be calculated on a single workstation.

Our method allows several marginally-stable eigenmodes to participate in the evolution of the background state simultaneously.
Consequently, we are able to exceed Herring's $\Ra = 10^6$ threshold.
For $\Ra > 10^6$ we did not observe any significant deviations from the trends observed in low-$\Ra$ MSTE.

MSTE contain large-scale convective cell structures, and have $\Nu\sim\Ra^{1/3}$.
These are similar to what has been found in experiments and direct numerical simulations.
But MSTE also exhibit unique and unexpected features: mean temperature gradient-reversals/dips, high kinetic energy flows, and a larger $\Nu$ than other equilibrated solutions. 
When initializing with different mean temperature profiles, we find the same MSTE, suggesting these equilibria might be unique.

Simulations initialized with the MSTE \eq{EQ:mste_ic} do not undergo an early convective transient period, but have faster flows when compared with DNS.
From a dynamical systems perspective, unstable orbits depart from MSTE and approach the global attractor on a viscous time scale.
This requires more computational effort to achieve relaxation when compared to the conventional conductive initial condition \eq{EQ:linear_ic}.
Initializing with only the temperature state of the MSTE avoids an early convective transient, but does not have excessive large-scale flows.
These states could be excellent initial conditions for high-Rayleigh number DNS.

Using the mean temperature in a statistically-steady DNS as a background state for an eigenvalue problem yields positive eigenvalues: the system is in a perpetual state of instability. 
Unstable modes tend to stabilize the system rapidly, creating a negative feedback loop whose average state is linearly unstable. 
Future work could adjust the marginal stability criterion to allow for moderately unstable modes.
Should the fast and slow time scales not be entirely separate, these modes might persist for long times. 
This might lead to greater agreement with DNS.

Another significant difference with DNS is the treatment of shear in the boundary layers.
At sufficiently high $\Ra$, the shear in the kinetic boundary layers could become unstable and turbulent \cite{Ahlers}.
This cannot occur in our quasilinear model because the background state has no velocity shear.
One avenue for future work is to include a mean horizontal flow, which could develop boundary layer structures.
In that case, we would search for states which also balance viscous and Reynolds stresses across the domain.
Perturbations about this background state could include shear flow instabilities.
Another approach would be to use the generalized quasilinear approximation, in which the background state includes both the mean and low-wavenumber temperature and velocity profiles \cite{marston2016}.
This would allow for linear instabilities of the shear driven by low-wavenumber rolls, as we see in all MSTE.
Generalized quasilinear calculations are inherently multi-dimensional, so they are more computationally expensive than the eigenvalue-based analysis used to find MSTE.
Despite these limitations, our one-dimensional MSTE states capture many interesting features of convection.

% To find MSTE, we initialize the time-evolution algorithm with the analytic temperature profile derived by \cite{Shishkina}.
% This involves modifying the boundary layer thickness $\delta_0$ to achieve marginal stability.

% Instead of imposing marginal stability, we can use quasilinear model to thermally-equilibrate and expand 1D approximations into 2D.
% As previously noted, there is no shortage of theories pertaining to the scaling behavior $\Nu \sim \Ra^{\beta}$ \cite{Malkus_1954, Howard_1966, Kraichnan, Spiegel, Castaing, Grossman, Ahlers}.
% For large $\Ra$, we can approximate $\delta_0 \approx \Nu^{-1}$ and construct a 1D temperature profile according to \eq{EQ:T0} or some other approximation.
% In this way, other 2D quasilinear thermal equilibria can be obtained and analyzed. 

\section*{Acknowledgments}
We would like to thank Charlie Doering, who greatly influenced the way many of us think about Rayleigh-B\'enard convection.
Some of the ideas in this work came from discussions with Charlie at Walsh Cottage during the WHOI GFD summer programs; it is difficult to imagining Walsh Cottage without Charlie's friendly and open scientific style, and enthusiasm for softball.
The authors thank Geoff Vasil, Greg Chini, Kyle Augustson, and Emma Kaufman for their valuable feedback and suggestions.
We would also like to thank Charlie Doering and Baole Wen for the tabulated DNS and ECS data for Figure~\ref{fig:nu_vs_ra}.
We thank the \texttt{Dedalus} and \texttt{Eigentools} development teams. 
Computations were conducted with support by the NASA High End Computing (HEC) Program through the NASA Advanced Supercomputing (NAS) Division at Ames Research Center on Pleiades with allocation GIDs s2276.

\appendix

\section{Initial Buoyancy Profile} \label{sec:initial_profile}
We initialize the thermal-equilibration algorithm with an analytical thermal boundary layer equation, derived by \cite{Shishkina} 
\begin{align}
    \overline{\mathcal{T}}_0(\xi) &= \frac{\sqrt{3}}{4\pi} \log \frac{(1 + a\xi)^3}{1 + (a\xi)^3} + \frac{3}{2\pi} \arctan \Big( \frac{4\pi}{9}\xi - \frac{1}{\sqrt{3}} \Big) + \frac{1}{4} \nonumber \\
    \xi &= \frac{z + 1/2}{\delta_0} \geq 0, \qquad a = \frac{2\pi}{3\sqrt{3}}\label{EQ:T0}
\end{align}
where $\delta_0$ is a new measure of the boundary layer height ($\delta_0 \neq \delta$ in general). 
This function is meant to describe the temperature near $z = -1/2$ so it does not pass through the origin.
An appropriate initial mean temperature profile $\overline{T}_0$ must be odd-symmetric, i.e.~$\overline{T}_0(z) = -\overline{T}_0(-z)$.
Due to continuity, this implies $\overline{T}_0(0) = 0$.
Accordingly, we construct $\overline{T}_0(z)$ by translating $\overline{\mathcal{T}}_0(z)$ vertically to pass through the origin.
We then take its odd-extension and include a scaling coefficient to satisfy the boundary conditions $\overline{T}_0(-1/2) = 1/2$ and $\overline{T}_0(1/2) = -1/2$.
The initial mean temperature profile is therefore given by
\begin{align*}
    \overline{T}_0(z) &= \frac{1}{2} \begin{cases}
        1 - \frac{\overline{\mathcal{T}}_0\big(\frac{z + 1/2}{\delta_0}\big)}{\overline{\mathcal{T}}_0\big(\frac{1}{2\delta_0}\big)} \quad -1/2 \leq z \leq 0 \\[0.5cm]
        -1 + \frac{\overline{\mathcal{T}}_0\big(\frac{1/2 - z}{\delta_0}\big)}{\overline{\mathcal{T}}_0\big(\frac{1}{2\delta_0}\big)} \quad 0 < z \leq 1/2.
    \end{cases}
\end{align*} 
 We expect each $\Ra$ to be associated with a unique $\delta_0$ for which $\overline{T}_0(z)$ is marginally-stable. 
It should be noted that when experimenting with various initial profiles $(\tanh, \, \rm{erf})$, we obtain indistinguishable equilibrated states.
Therefore these initial states lie in the MSTE basin of attraction with respect to the quasilinear system.
This might also suggest that solutions are unique. 
An example of \eq{EQ:T0} is given by the blue curve in Figure~\ref{fig:T0_profiles}.

\section{Timestep Management} \label{sec:timestep}
To find MSTE, we advance $\overline{T}$ by timesteps of size $\Delta t$.
For large $\Delta t$, we find the advective flux terms fail to provide a stabilizing influence.
This is due to coupling between the eigenfunctions $W(z)$, $\theta(z)$ and the mean temperature profile $\overline{T}$.
In the context of our algorithm, this effectively deletes the sought-after root of the maximum eigenvalue function $\sigma_{\rm{max}}(A^2)$, causing the root-finding method to fail. 
This is not a numerical instability, rather, it is an inherent limitation of our timestepping algorithm.
To curtail this, we halt the root-finding algorithms after 20 successive approximations, and reduce the timestep by a factor of 1.1, and try again.

The timestep must be also reduced to avoid a numerical instability.
We find that for large $\Delta t$, after several hundred iterations, highly concave features develop in the advective flux term $\langle w'T' \rangle_x$ near $z=0$. 
Such features are undesired, as they are uncommon in similar calculations \cite{Malkus_1954} and we do not believe they accurately represent a physical process.
%Further, we expect our algorithm to converge to the analytic solution of \eq{EQ:T0_IVP} for small $\Delta t$.
%Therefore results obtained uniquely via the use of relatively large $\Delta t$ must not agree with the analytic solution.
If ignored, the concave features grow in magnitude until they affect $\overline{T}$ on a readily apparent scale.
Eventually $\overline{T}$ develops oscillations near $z = 0$ and the timestep must be reduced.
Once these oscillations reach some amplitude, they cannot be eliminated via timestep reduction and the roots of $\sigma_{\rm{max}}(A^2)$ vanish.
To avoid this, we measure the curvature of the advective flux $|\frac{\partial^2}{\partial z^2} \langle w'T' \rangle_x|_{z=0}|$ of the dominant $k_x = 1.5\pi$ mode.
We find that reducing the timestep $\Delta t$ by the same factor of 1.1 whenever this curvature measure exceeds $10^{-6}$ curtails the problem.
Both of these issues appear to become more prominent as the boundary layers diffuse over the course of the algorithm's execution (see Figure~\ref{fig:T0_profiles}).
Thus there is never a practical opportunity to increase the timestep.

\onecolumngrid
\section{MSTE Parameters and Results}\label{sec:tables}

\begin{longtable*}{ccccl}
    %\begin{tabular}{cccccc}
    %\hline
    %\begin{longtable}[c]{@{}*{6}{>{\arraybackslash}p{0.15\linewidth}}@{}}
    \quad\quad\quad $\Ra$	\quad\quad\quad	&	\quad\quad\quad		$N_z$	\quad\quad\quad		&	\quad\quad\quad		$\Nu$	\quad\quad\quad		&	\quad\quad\quad		$\delta$	\quad\quad\quad		&	\quad\quad	marginal $k_x$	\quad\quad\quad\\[0.03cm]
    \hline
    $10^5$ & 256   & 5.9343  & 0.15746 \;    &   \; 1$\pi$, 1.5$\pi$   \\[0.03cm]
    $2 \times 10^5$ & 256   & 7.45986 & 0.12341 \;    &   \; 1.5$\pi$   \\[0.03cm]
    $4 \times 10^5$ & 256   & 8.92597 & 0.10395 \;    &   \; 1.5$\pi$   \\[0.03cm]
    $10^6$ & 256   & 11.6689 & 0.07922 \;    &   \; 1.5$\pi$, 2.5$\pi$, 3$\pi$   \\[0.03cm]
    $2 \times 10^6$ & 256   & 14.8463 & 0.06345 \;    &   \; 1.5$\pi$, 3.5$\pi$   \\[0.03cm]
    $4 \times 10^6$ & 256   & 18.9433 & 0.04933 \;    &   \; 1.5$\pi$, 4.5$\pi$   \\[0.03cm]
    $10^7$ & 256   & 25.6821 & 0.03632 \;    &   \; 1.5$\pi$, 5.5$\pi$   \\[0.03cm]
    $2 \times 10^7$ & 256   & 32.4531 & 0.02820 \;    &   \; 1.5$\pi$, 6.5$\pi$, 7$\pi$  \\[0.03cm]
    $4 \times 10^7$ & 512   & 40.7925 & 0.02289 \;    &   \; 1.5$\pi$, 8$\pi$, 8.5$\pi$   \\[0.03cm]
    $10^8$ & 512   & 55.4383 & 0.01690 \;    &   \; 1.5$\pi$, 4.5$\pi$, 10.5$\pi$, 11$\pi$  \\[0.03cm]
    $2 \times 10^8$ & 512   & 69.8349 & 0.01318 \;    &   \; 1.5$\pi$, 5$\pi$, 13.5$\pi$, 14$\pi$   \\[0.03cm]
    $4 \times 10^8$ & 512   & 87.8525 & 0.01053 \;    &   \; 1.5$\pi$, 5.5$\pi$,17.5$\pi$   \\[0.03cm]
    $10^9$ & 768   & 119.318 & 0.00817 \;    &   \; 1.5$\pi$, 6.0$\pi$, 6.5$\pi$, 23.5$\pi$, 24$\pi$ \\\vspace{0.1in}\\
    %\hline
    %  \end{tabular}
    \caption{Control parameters and results are given for the MSTE timestepping algorithm. $N_z$ denotes the number of Chebyshev basis functions employed. 
    This, along with $\Ra$ are specified on initialization.
    The remaining quantities are computed directly from the MSTE. 
    The low $\Ra = 10^5$ case requires $\sim 12$ hours to compute while the high $\Ra = 10^9$ case requires $\sim 72$ hours. 
    Eigenvalue solves are performed for various $k_x$ simultaneously using 28 cores.}
    {\label{tab:metrics}}
    \end{longtable*}

\section{MSTE Eigenfunctions}\label{sec:eigenfunctions}

\begin{figure}[H]
    \centering
    \includegraphics[width=6.75in]{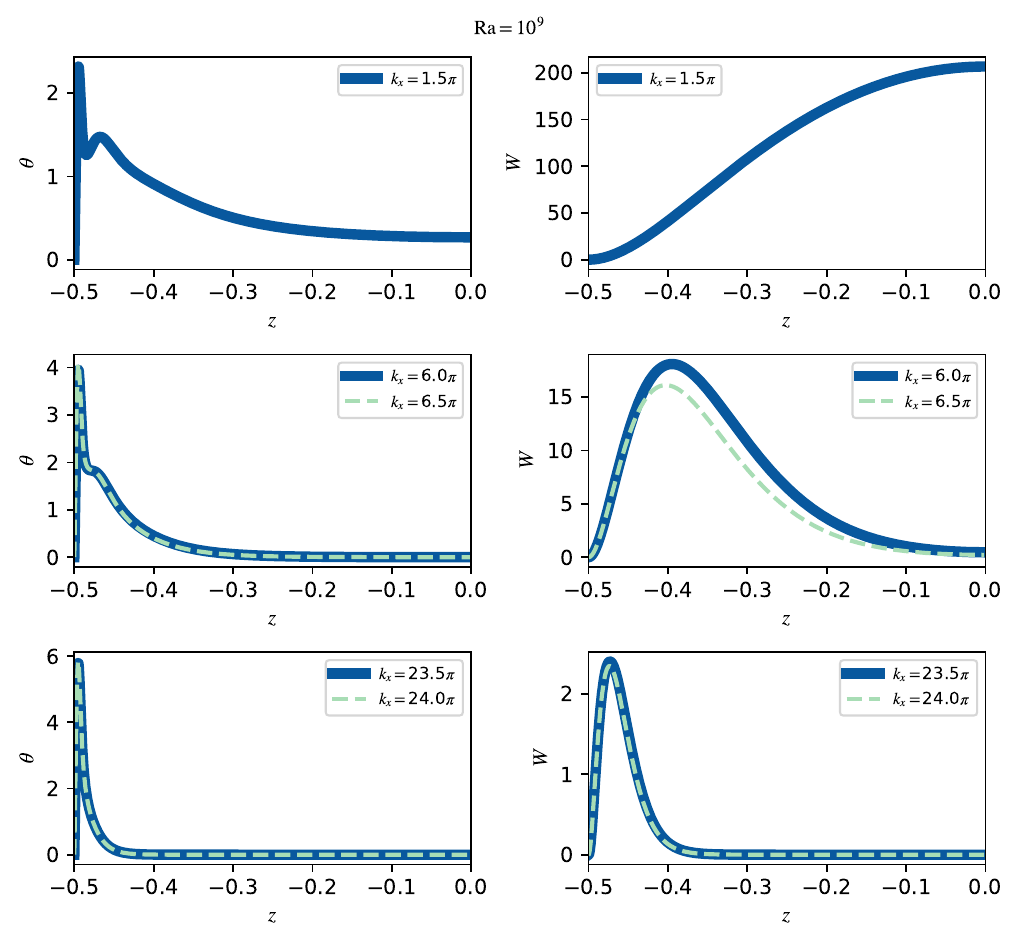}
    \caption{The temperature and pressure MSTE eigenfunctions are shown in separate plots for various marginally-stable modes. 
    The eigenfunctions are normalized such that $\langle |\theta|^2\rangle_{\mathcal{D}}=1$.
    We select a phase such that the eigenfunctions are real.
    In this case the eigenfunctions are even which we exploit by plotting over half the domain.
    }%
    \label{fig:eigenfunctions}%
\end{figure}

In Figure~\ref{fig:eigenfunctions} we plot the velocity and temperature perturbation eigenfunctions for the marginally-stable modes in the $\Ra =10^9$ MSTE.
Though the eigenfunctions are in general complex, the linear system \eqss{EQ:linear1}{EQ:linear3} obeys phase-shift symmetry.
Any set of solutions can be phase-shifted by multiplying the perturbations by a complex quantity of unit modulus $e^{i\phi}$.
In this particular case, we can eliminate the perturbations' imaginary components by selecting the proper $\phi$.
This is due to the apparent exchange of stabilities observed previously: $\omega = 0$ at marginal stability implies that the system admits a set of real solutions.
Adjacent modes, whose wavenumbers only differ by $0.5\pi$, are superimposed to illustrate their similarity.
For the large-scale $k_x = 1.5\pi$ mode, we find the vertical velocity eigenfunction $W$ is maximized near the center of the domain.
As $k_x$ increases, we observe two vertical velocity maxima, each tending towards their respective boundary.
However, in all cases, the temperature perturbations are more concentrated at the boundaries than the vertical velocity.
This trend is also apparent in the two pairs of adjacent modes.
In the temperature eigenfunctions $\theta$, small-scale zig-zags appear near the boundaries at $k_x = 1.5\pi$, subsiding as $k_x$ increases.
The left-most peaks of these temperature perturbation profiles approximately coincide with the previously-defined boundary layer height $\delta$.

\twocolumngrid
\clearpage

\bibliography{ms_rbc.bib}

\end{document}